\documentclass[%
reprint,
groupedaddress,
openany
bibnotes,
 amsmath,amssymb,
 aps,
prc,
]{revtex4-1}

\usepackage{graphicx}
\usepackage{epstopdf}
\usepackage{dcolumn}
\usepackage{bm}

\begin{document}

\preprint{APS/123-QED}

\title{Multi-alpha Boson Gas state in Fusion Evaporation Reaction and Three-body Force}
\email{tfwang@buaa.edu.cn (T. F. Wang)}
 \author{Taofeng Wang$^{1*}$, Ziming Li$^{1}$, R. B. Wiringa$^{2}$, Minliang Liu$^{3}$, Jiansong Wang$^{3}$,\\ Yanyun Yang$^{3}$, Qinghua He$^{4}$, Zhiyu Sun$^{3}$, Chengjian Lin$^{5}$, M. Assié$^{6}$,Y. Ayyad$^{7}$,\\ D. Beaumel$^{6}$, Zhen Bai$^{3}$, Fangfang Duan$^{3}$, Zhihao Gao$^{3}$, Song Guo$^{3}$, Yue Hu$^{1}$,\\ Wei Jiang$^{8}$,  F. Kobayashi$^{9}$, Chengui Lu$^{3}$, Junbing Ma$^{3}$, Peng Ma$^{3}$,\\ P. Napolitani$^{10}$, G. Verde$^{11,12}$, Jianguo Wang$^{3}$, Xianglun Wei$^{3}$, Guoqing Xiao$^{3}$,\\ Hushan Xu$^{3}$, Biao Yang$^{8}$, Runhe Yang$^{3}$, Yongjin Yao$^{1}$, Chaoyue Yu$^{3}$,\\ Junwei Zhang$^{3}$, Xing Zhang$^{3}$, Yuhu Zhang$^{3}$,  Xiaohong Zhou$^{3}$}
\affiliation{%
 $^{1}$School of Physics, Beihang University, Beijing 100191, China\\ 
 $^{2}$Physics Division, Argonne National Laboratory, Argonne, Illinois 60439, USA\\
 $^{3}$Institute of Modern Physics, Chinese Academy of Sciences, Lanzhou 730000, China\\
 $^{4}$Department of Nuclear Science $\&$ Engineering, College of Material Science and Technology,\\Nanjing University of Aeronautics and Astronautics, Nanjing 210016, China\\ 
 $^{5}$China Institute of Atomic Energy, P.O. Box 275 (10), Beijing 102413, China\\
 $^{6}$IJCLab, Université Paris-Saclay, CNRS/IN2P3, 91405 Orsay, France\\
 $^{7}$Facility for Rare Isotope Beams, Michigan State University, East Lansing, Michigan 48824, USA\\
 $^{8}$State Key Laboratory of Nuclear Physics and Technology, School of Physics, Peking University, Beijing 100871, China\\
 $^{9}$Graduate School of Engineering Science, Osaka University, 1-3 Machikaneyama, Toyonaka, Osaka 560-8531, Japan\\
 $^{10}$IPN, CNRS/IN2P3, Université Paris-Sud 11, Université Paris-Saclay, 91406 Orsay Cedex, France\\
 $^{11}$INFN Sezione di Catania, via Santa Sofia 64, I-95123 Catania, Italy\\
 $^{12}$Laboratoire des 2 Infinis - Toulouse (L2IT-IN2P3), Université de Toulouse, CNRS, UPS, F-31062 Toulouse Cedex 9, France
  }

\begin{abstract}
The experimental evidence for the $\alpha$ Boson gas state in the $^{11}$C+$^{12}$C$\rightarrow$$^{23}$Mg$^{\ast}$ fusion evaporation reaction is presented. By measuring the $\alpha$ emission spectrum with multiplicity 2 and 3, we provide insight into the existence of a three-body force among $\alpha$ particles. The observed spectrum exhibited distinct tails corresponding to $\alpha$ particles emitted in pairs and triplets consistent well with the model-calculations of AV18-UX and chiral effective field theory of NV2-3-la*, indicating the formation of $\alpha$ clusters with three-body force in the Boson gas state. 


\end{abstract}

\pacs{Valid PACS appear here}
\maketitle


\section{Introduction}
Fusion evaporation is a fascinating phenomenon that occurs when two atomic nuclei collide and merge to form a compound nucleus, which subsequently undergoes a de-excitation process by emitting particles. This process can be analyzed and understood from various aspects, including the excitation energy, spin of the compound nucleus and multiplicity of the emitted identical particles [1-8]. By studying the emission patterns of particles and their correlations with these parameters, ones can gain insights into the nuclear structure, reaction mechanisms, and the properties of compound nuclei. This knowledge is crucial for various fields, including nuclear astrophysics, nuclear energy production, and the synthesis of superheavy elements. 

The excitation energy determines the available energy for particle emission and governs the probability and types of particles emitted during the de-excitation process. The angular momentum associated with the colliding nuclei is transferred to the compound nucleus upon fusion. The spin of the compound nucleus influences the selection rules for subsequent particle emission. The conservation of angular momentum dictates the possible spin states of the emitted particles and their relative intensities. The spin distribution of the compound nucleus is crucial in determining the probabilities and types of particles emitted during the de-excitation process. The multiplicity of emitted particles is determined by the excitation energy and spin of the compound nucleus. Higher excitation energies and larger spin values generally lead to higher multiplicities. 

The $\alpha$ particle Boson gas dilute state [9, 10] refers to a specific condition in fusion evaporation reactions where the $\alpha$ particles behave like a gas of non-interacting particles. This approximation is particularly applicable when the density of $\alpha$ particles within the colliding nuclei is low, and the inter-alpha particle interactions can be neglected. The $\alpha$ particle Boson gas model provides a useful framework for understanding the initial stages of the reaction and the subsequent evaporation process involving $\alpha$ particles. This is typically the case in heavy-ion reactions involving light nuclei or when the incident energy is sufficiently high to overcome the nuclear interaction barriers. Under these conditions, the probability of inter-alpha particle collisions becomes small enough that the $\alpha$ particles can be treated as quasi-independent particles within the Boson gas model.

The Boson gas model assumes that the $\alpha$ particles within the colliding nuclei follow a statistical distribution known as the Bose-Einstein distribution [9], which characterizes the occupancy of energy levels within the system. The Bose-Einstein distribution determines the average energy, momentum, and occupation probability of the particles within the Boson gas [9]. Within the $\alpha$ particle Boson gas model, the dilute state assumption allows for a simplified treatment of the collision dynamics and subsequent de-excitation process involving $\alpha$ particles emission. 

The low $\alpha$ particle density and negligible inter-alpha particle interactions enable a decoupling of the collision and evaporation stages, simplifying the analysis. During the fusion stage, when the colliding nuclei approach each other, the attractive nuclear force acts on the nucleons individually, including the $\alpha$ particles. As a result, the $\alpha$ particles are treated as independent entities during the fusion process. Following the formation of the compound nucleus, which consists of the fused nuclei and any emitted particles, it rapidly equilibrates internally through various nucleon-nucleon interactions, allowing it to reach a high excitation energy. The compound nucleus then undergoes a de-excitation process, where it emits particles, including $\alpha$ particles, until it reaches a more stable state. 

The three-body force among $\alpha$ particles [11-21] refers to the additional nuclear force that arises from the interactions among multiple $\alpha$ particles in a system. It accounts for the collective interaction between the constituent nucleons of the $\alpha$ particles, extending beyond the pairwise interactions. When multiple $\alpha$ particles come close to each other, their mutual interactions give rise to a three-body force that cannot be fully explained by the pairwise interactions alone. 

The three-body force affects various aspects of systems involving multiple $\alpha$ particles. The additional attraction arising from the three-body force enhances the stability of $\alpha$ clusters, contributing significantly to their binding energy. Furthermore, the three-body force influences the spatial arrangement and structure of systems involving multiple $\alpha$ particles. The interplay between the pairwise interactions and the three-body force determines the overall spatial arrangement and clustering patterns of $\alpha$ particles. 

\section{Experimental procedure}
The present experimental measurement was performed at the Radioactive Ion Beam Line at the Heavy Ion Research Facility in Lanzhou (HIRFL-RIBLL) [22], as shown in Fig. 1. A 60 MeV/nucleon $^{12}$C beam was transfed to bombard a 3.5 mm $^{9}$Be target to produce about 25 MeV/nucleon $^{11}$C secondary beam with a purity of about 99$\%$ and an intensity of about 10$^{4}$ particles per second [23]. The beam particles were identified in terms of $B\rho-$TOF$-\Delta E$ method with the magnets and two plastic scintillator detectors in the beam line [22]. The $^{11}$C secondary beam were bombarded on a 50 mg/cm$^{2}$ carbon target to produce the breakup reaction. 

\begin{figure}[htb]
\includegraphics[width=8.5cm]{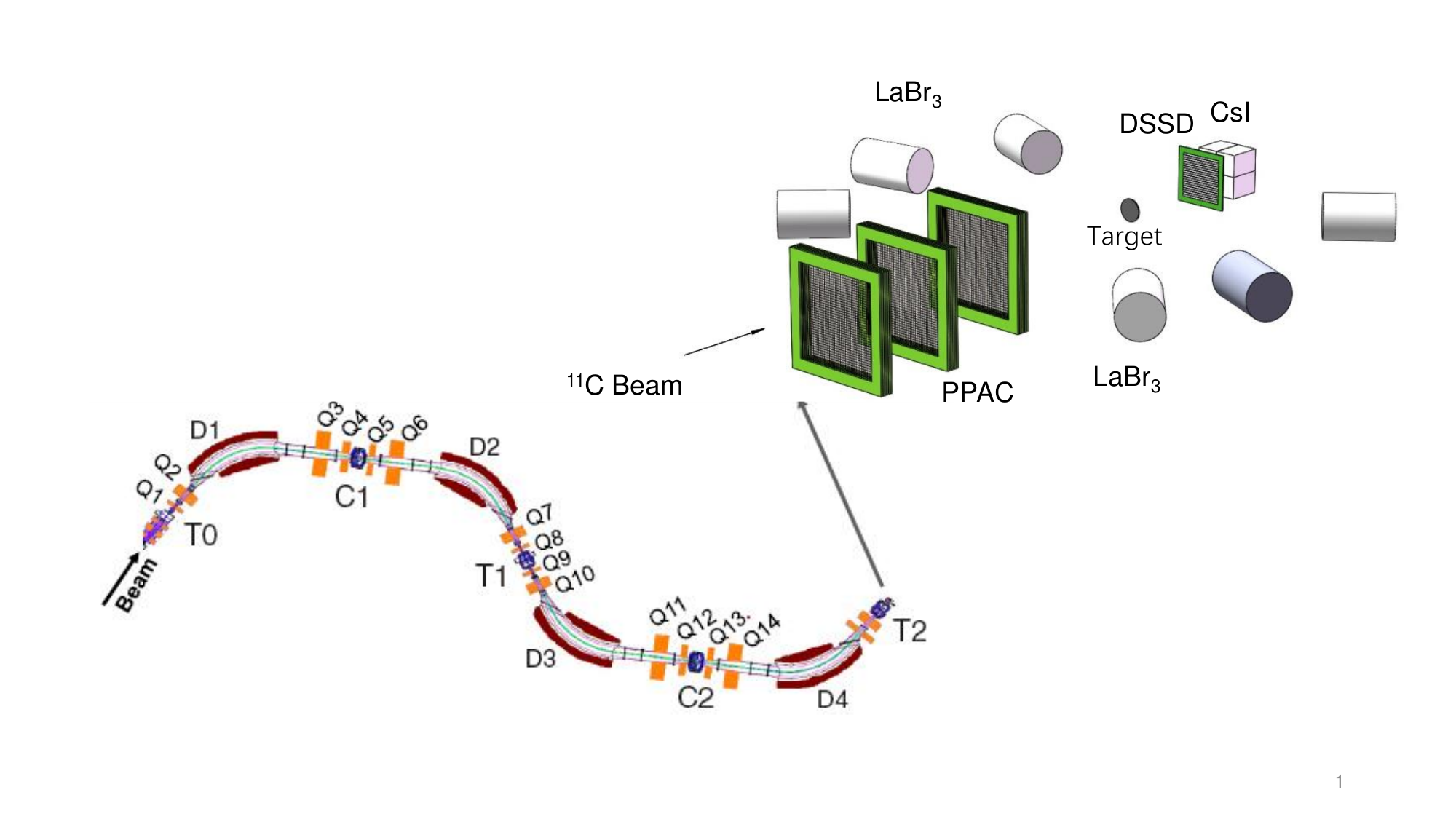}
\caption{\label{fig:pdfart} The experimental equipment consisting of three PPACs for determing the reaction position of beam particle on the target, the reaction products were detected by a DSSD silicon detector combined with a 2$\times$2 CsI(Tl) scintillators array, the decay $\gamma$s were detected by five LaBr$_{3}$ (Ce) and one NaI scintillator detectors.}
\end{figure}

Three parallel plate avalanche chambers (PPACs) with 50$\times$50 mm$^{2}$ active area and position resolutions of about 1 mm (FWHM) in both the $X$ and $Y$ directions were placed in front of the target to track the incident $^{11}$C beam [22] and to subsequently get the reaction vertex in the target. $\alpha$ particles are detected by the zero-degree telescope system which consists of a double-sided silicon strip detector (DSSD, of 148 $\mu$m in thickness and 50$\times$50 mm$^{2}$ in cross-sectional area) with 32 strips on both front and back sides, and a 2$\times$2 photodiode (PD) readout CsI (Tl) scintillator (25$\times$25$\times$30 mm$^{3}$ size for each unit) array. Each CsI (Tl) scintillator is covered by two layers of high reflection Tyvek papers and a 10 $\mu$m aluminum coated Mylar film as window. The PD is coupled to CsI (Tl) scintillator with the photoconductive silicone grease. The angular coverage of the zero-degree telescope is about 0-9$^{\circ}$. Five LaBr$_{3}$ (Ce) and one NaI scintillator detectors were placed around the target to measure the decayed $\gamma$s from the excited fragments. DSSD was utilized to record the $\Delta$E energy and the position of the detecting fragments, therefore, the emission angle may be obtained by combining with the reaction vertex in the target. CsI (Tl) detection system provides the residual E energy of the fragments. Particle identifications (PID) for $\alpha$s were performed using $\Delta$E-E contour, as shown in Fig. 2. The energy resolution with sigma of this  $\Delta$E-E detection system is estimated to be $\sim$0.8 MeV from numerical simulation.

\begin{figure}[htb]
\includegraphics[width=9.2cm]{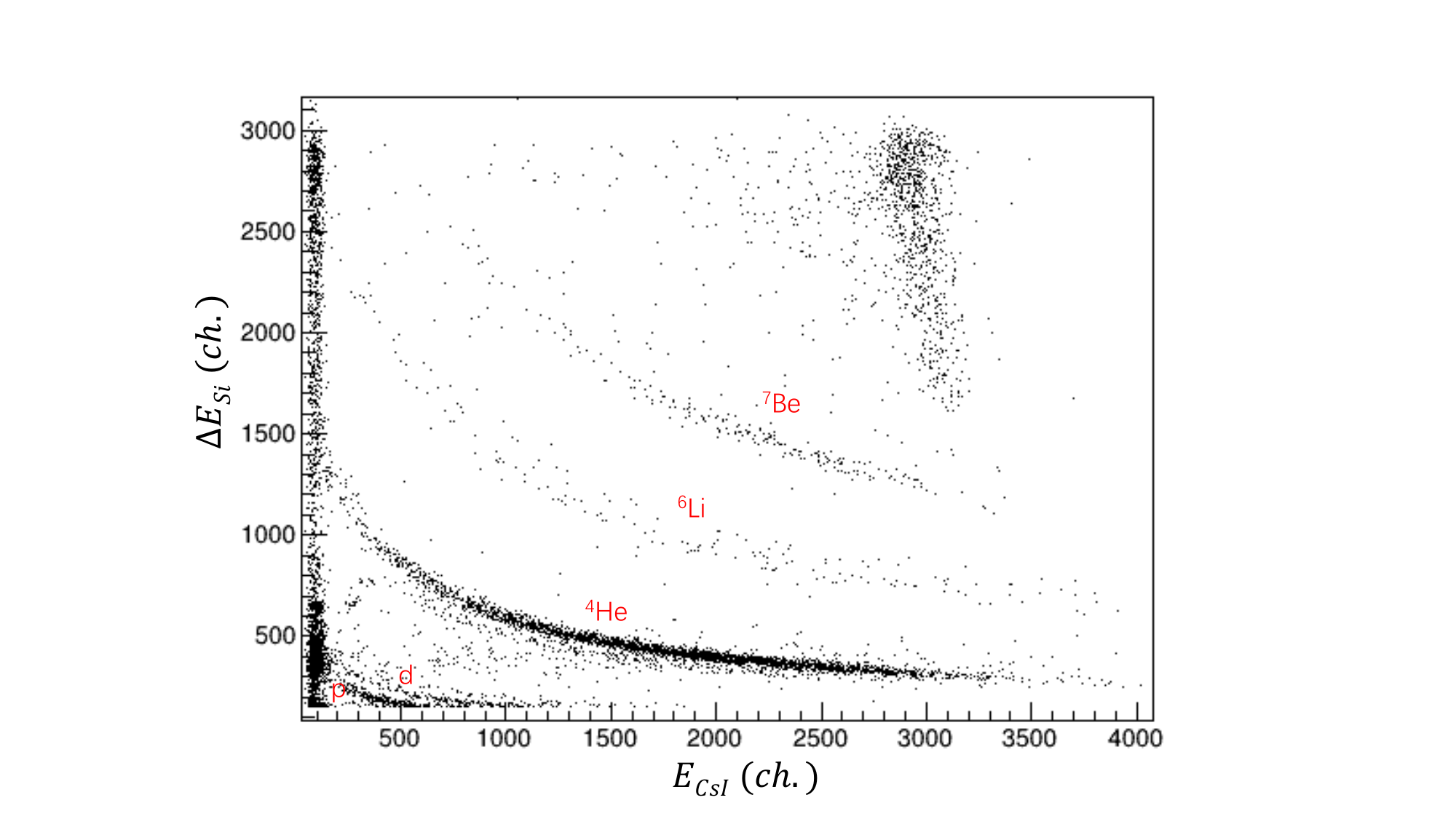}
\caption{\label{fig:pdfart}Particle identification (PID) with $\Delta E$ strip-silicon detector and $E$ CsI array.}
\end{figure}

\section{ Data analysis procedure}
\subsection{Analysis with statistical model and chiral effective field theory}

The $\alpha$ spectra with the multiplicity equal to 2 and 3 depending on $p_{c.m.}$ the $\alpha$ momentum in the center of mass system were sorted and exhibitted in Fig. 3, in which the center value of the maxmium amplititude with $p_{c.m.}$ momentum of 3$\alpha$ spectrum moves to the higher compare to that of 2$\alpha$. This aspect reflects the higher local temperature of 3$\alpha$ according to Maxwell-Boltzmann theory of the root mean square speed $\sqrt{\overline{v}^{2}}$$\approx$1.73$\sqrt{RT/M_{\alpha}}$, where $R$ is the universal gas constant, $T$ is the gas temperature, $M_{\alpha}$ is the mass of $\alpha$ particle.

The absolute slope value of the spectrum tail of 2$\alpha$ is larger than that of 3$\alpha$ as shown in Fig. 3. The formation of such cluster configurations can be understood in terms of a loosely multi-$\alpha$ bound system [24, 25]. Assuming a square well potential between them, the density distribution of $\alpha$ is $\rho (r)=|\Psi(r)|^{2}\propto \frac{e^{-2\kappa r}}{r^{2}}$, where $\Psi(r)$ is the $\alpha$ particle wave function. Its slope of the density distribution tail in $r$ position coordinate space is determined by the quantity $\kappa$. The momentum distribution of $\alpha$ obtained from the Fourier transform for wave function $\Psi(r)$ can be expressed by $f(p)=C/(p_{i}^{2}+\kappa^{2})$ in the momentum space. $\kappa_{2\alpha}$$>$$\kappa_{3\alpha}$ is observed according to $f(p)$ from the comparison of the the slopes of $3\alpha$ and $2\alpha$ in Fig. 3, it hints the long tail distribution of $3\alpha$ in position coordinate space.

\begin{figure}[htb]
\includegraphics[width=9.2cm]{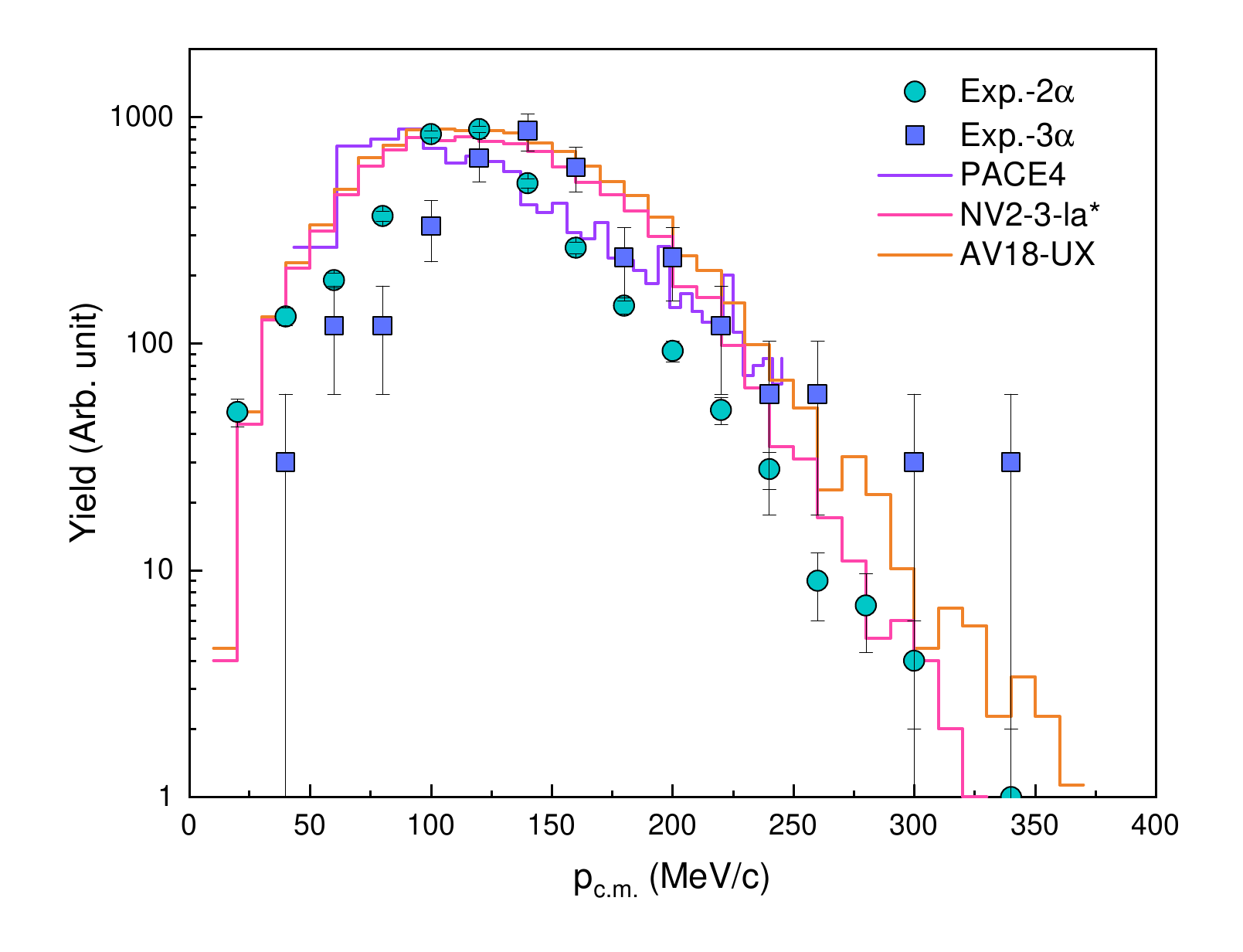}
\caption{\label{fig:pdfart}The spectra of 2$\alpha$ and 3$\alpha$ compared with the statistical mode of PACE4 calculation [28] as well as the calculations with AV18-UX potential combined by three-body force [26], besides togethers with the calculation with chiral effective field theory of NV2-3-la* model [27].}
\end{figure}

The fusion evaporation process begins with the collision of two atomic nuclei with enough energy to overcome the repulsive Coulomb barrier between them. This collision leads to the formation of a compound nucleus, which is a temporary, highly excited state of the combined system. Upon forming the compound nucleus, the incident kinetic energy is converted into internal energy, causing the compound nucleus to become even more excited. The excitation energy is distributed among the nucleons within the nucleus, resulting in the collective motion of the nucleons, such as vibrations and rotations.

In the fusion evaporation process, the formation of a compound nucleus involves the interaction between the colliding atomic nuclei and the subsequent rearrangement of nucleons within the compound nucleus. This rearrangement results in the establishment of a mean field and the presence of local $\alpha$ two-body and three-body interactions. The formed mean field refers to the average potential experienced by the nucleons within the compound nucleus. It arises due to the collective interaction of all nucleons and is influenced by the nuclear structure and the density distribution of nucleons. The mean field is responsible for governing the motion and behavior of individual nucleons within the compound nucleus. 

In addition to the mean field, the fusion evaporation process also involves local $\alpha$ two-body and three-body interactions. These interactions arise due to the formation and subsequent decay of $\alpha$ particles within the compound nucleus. This three-body interaction can influence the energies and angular momenta of the nucleons involved and can influence the decay channels and final states of the compound nucleus. 

The local $\alpha$ two-body and three-body interactions are typically incorporated into theoretical models and simulations of the fusion evaporation process. These interactions are often described using effective nuclear potentials or effective field theories, which aim to capture the essential features of the $\alpha$-particle interactions within the compound nucleus. In Fig. 3 the calculations of the adopted Aagonne v18 two-nucleon potential and Urbana X three-nucleon potentials [26] as well as the calculations with Nv2-3-la$^{\ast}$ chiral effective field theory ($\chi$EFT) [27] including three-nucleon force are consistent well to the tail slope of 3$\alpha$. The fusion evaporation statistical model PAEC4 [28] for the calculation are partly consistent to the tail slopes of 2$\alpha$ and 3$\alpha$ spectra. The three-body interaction among $\alpha$ particles arises from the complex interplay between the nucleons within each $\alpha$ particle and the collective interactions among the $\alpha$ particles themselves. The three-body force cannot be fully explained by the pairwise interactions alone and requires a more advanced theoretical treatment to capture its effects accurately. $\chi$EFT has been utilized to study the three-body force among alpha particles. It combines the principles of chiral symmetry and effective field theory and provides a systematic expansion for nuclear interactions, including the three-body force. By applying $\chi$EFT, ones have been able to derive interactions among alpha particles that accurately reproduce experimental data and provide valuable insights into the three-body force. These aspects indicate that the effects of three-body interaction among $\alpha$ particles prefer to the formation of multiple $\alpha$ particles dilute state in the final fusion process of $^{11}$C+$^{12}$C$\rightarrow$$^{23}$Mg$^{\ast}$ reaction.

\begin{figure}[htb]
\includegraphics[width=9.5cm]{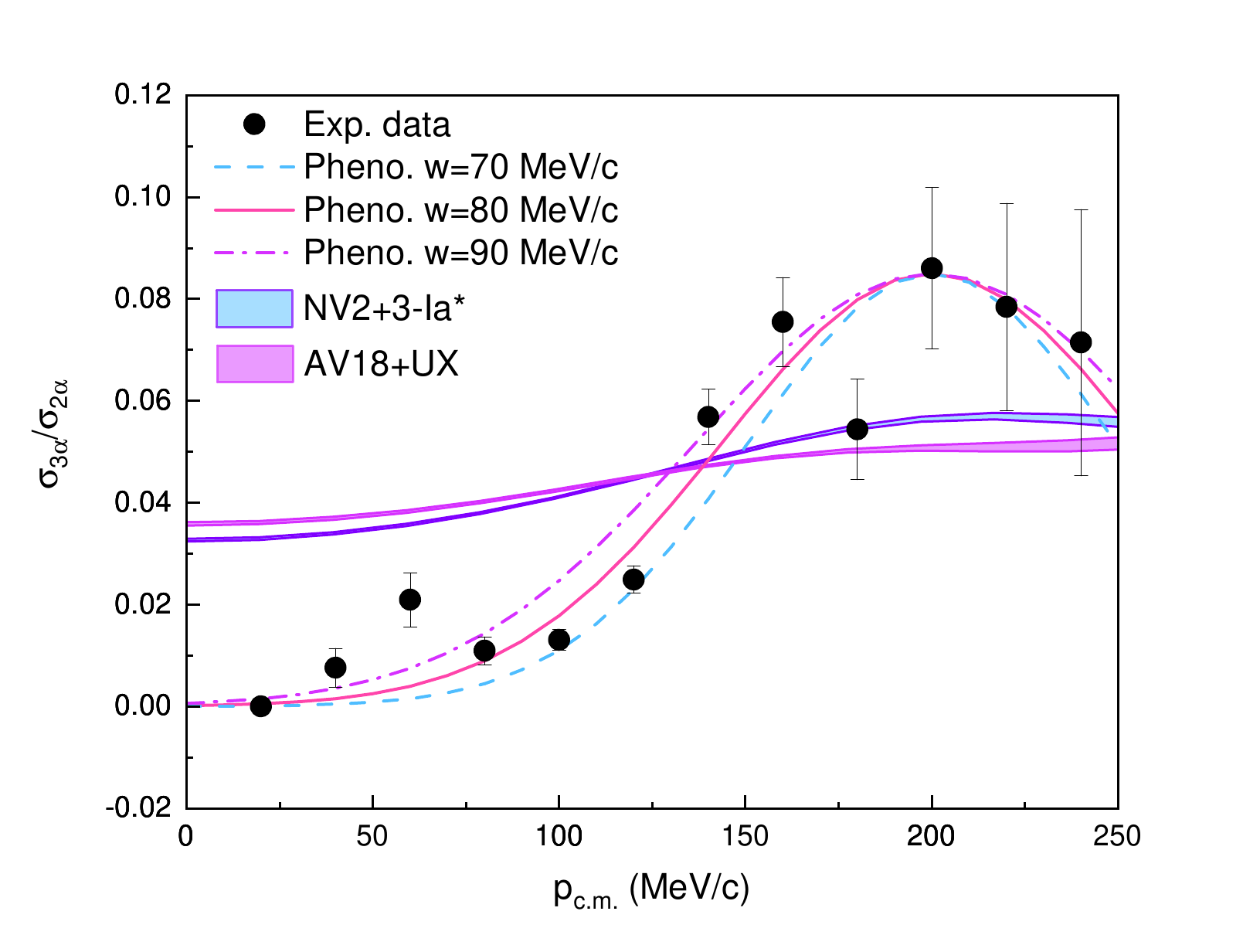}
\caption{\label{fig:pdfart}The cross section ratio of $\sigma_{3\alpha}/\sigma_{2\alpha}$ dependent on the $\alpha$ momentum $p_{c.m.}$, together with the calculations of NV2+3+la* and AV18+UX models, besides a phenomenological curves with different widths reflect $\alpha$ three-body force dependent on $p_{c.m.}$. }
\end{figure}

\subsection{Enhancement of three-body force with $p_{c.m.}$}
In fusion evaporation reactions, the cross section ratio of the 3$\alpha$ spectrum to the 2$\alpha$ spectrum can provide valuable information about the dynamics and properties of the reaction. This ratio of $\sigma_{3\alpha}/\sigma_{2\alpha}$ is dependent on the $\alpha$ momentum $p_{c.m.}$ and exhibits an increasing trend as shown in Fig. 4. The cross section ratio of the 3$\alpha$ spectrum to the 2$\alpha$ spectrum refers to the relative number of events or probabilities of observing three $\alpha$ particles versus two $\alpha$ particles in the final state of a fusion evaporation reaction. The dependence of this ratio on the $\alpha$ momentum $p_{c.m.}$ arises from the underlying dynamics of the fusion evaporation process. As the $\alpha$ momentum increases, the kinetic energy available for the system also increases. This higher kinetic energy allows for more energetic collisions and greater excitation of the compound nucleus formed during the reaction.

The strong nucleon-nucleon interaction was formulated by the Yukawa theory of multi-meson exchange in the short distance, and one pion meson exchange in the large distance case. Two nucleon One Pion Exchange Potential (OPEP)[29] was described by

\begin{equation}
\begin{split}
V^{1\pi}_{2N}=g^{2}\frac{m_{\pi}}{12M^{2}\hbar}\boldsymbol{\tau}(1)\cdot\boldsymbol{\tau}(2)[\boldsymbol{\sigma}_{1}\cdot\boldsymbol{\sigma}_{2}+S_{12}(1+\frac{2}{x}+\frac{3}{x^{2}})\frac{e^{-x}}{x}],
\end{split}
\end{equation}

where $x=\frac{m_{\pi}c}{\hbar}r$, $r$ is the distance of two particles, $M=\frac{1}{2}(m_{p}+m_{n})$ is the reduced mass of nucleon, $\sigma_{i}$($\tau_{i}$) is the Cartesian component of spin(isospin) operators of nucleon $i$, $S_{12}$ denotes the non-central interaction of tensor force term. On the other hand, three-body one pion exchange potential was shown as [11, 30]

\begin{equation}
\begin{split}
V^{1\pi}_{3N}=-\frac{g_{A}}{8f^{2}_{\pi}}\frac{c_{D}}{f^{2}_{\pi}\Lambda_{\chi}}\sum_{i\neq j\neq k}\frac{\boldsymbol{\sigma}_{j}\cdot \boldsymbol{Q}_{j}}{Q^{2}_{j}+m^{2}_{\pi}}(\boldsymbol{\tau}_{i}\cdot\boldsymbol{\tau}_{j})(\boldsymbol{\sigma}_{i}\cdot \boldsymbol{Q}_{j}),
\end{split}
\end{equation}

where $\boldsymbol{Q}_{j}=\boldsymbol{k}^{'}_{i}-\boldsymbol{k}_{i}$ is the transfer momentum, i.e. the difference between the initial and final single-particle momenta ($\boldsymbol{k}^{'}_{i}$ and $\boldsymbol{k}_{i}$ respectively). Since the interactions among $\alpha$ particles origiante from the residule interactions of two-nucleon force and three-nucleon force within $\alpha$ particle, the mutual $\alpha$ interactions still can be expressed by two-body and three-body force with Eq. (1) and (2), named by $V^{1\pi}_{2\alpha}$ and $V^{1\pi}_{3\alpha}$. $V^{1\pi}_{2\alpha}$ is position $r$ dependence, while $V^{1\pi}_{3\alpha}$ is momentum $Q_{j}$ dependence. Thereby, the potential ratio of $V^{1\pi}_{3\alpha}$ to $V^{1\pi}_{2\alpha}$ is momentum $Q_{j}$ positive dependence that leads to the enhancement of three-body force effects at high momentum transfer, it is clearly reflected by the plots of $\sigma_{3\alpha}/\sigma_{2\alpha}$ with $p_{c.m.}$ in Fig. 4. 

A phenomenological calculation has been parametrized in terms of the difference between single-particle momentum and Fermi momentum $p_{F}$ and distribution width. As shown in Fig. 4, the maximum of the increase ratio of $\sigma_{3\alpha}/\sigma_{2\alpha}$ around Fermi momentum $p_{F}$ indicates the three-body force is favorable of high momentum admixture.  

\begin{equation}
\begin{split}
\sigma_{3\alpha}/\sigma_{2\alpha}=Ce^{-\frac{(p-p_{F})^{2}}{w^{2}}},
\end{split}
\end{equation}

where $p$ equal to $\alpha$ momentum $p_{c.m.}$, $C$=0.085, the Fermi momentum $p_{F}$=200 MeV/c, $w$=70, 80, 90 MeV/c, these plots are consistent well to the cross section ratio $\sigma_{3\alpha}/\sigma_{2\alpha}$  of this measurement. These phenomenological calculations indicate that the momentum dependent three-body force among multi-alpha particles are essentially enhanced in high momentum region, especially near Fermi momentum. The fevorable $\alpha$ dilute Boson gas state, thereby, is formed in the final process of fusion reaction.  

\begin{figure}[htb]
\includegraphics[width=9.2cm]{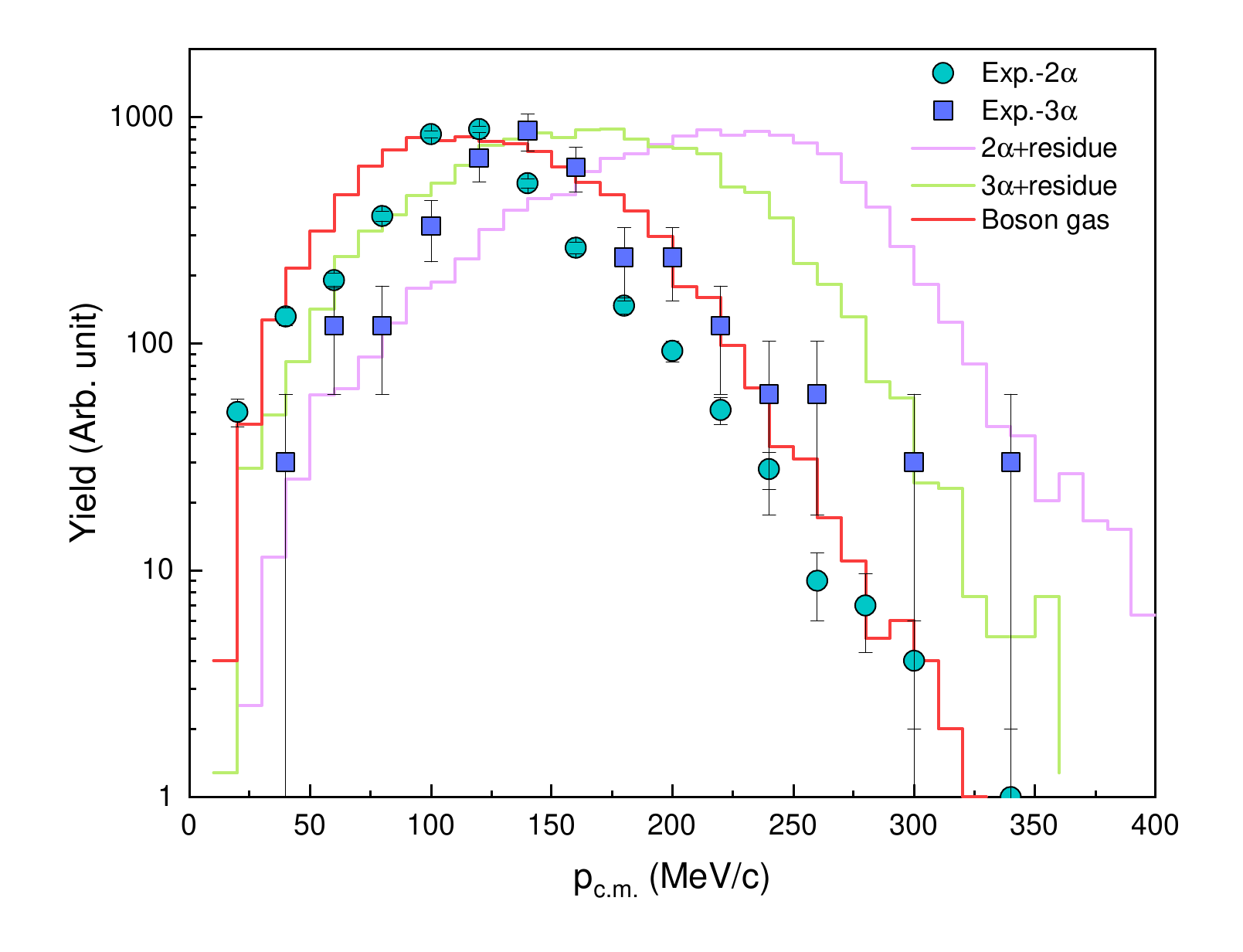}
\caption{\label{fig:pdfart}The spectra of 2$\alpha$ and 3$\alpha$ compared with the models of 2$\alpha$+residule nucleus, 3$\alpha$+residule nucleus and all of Helium Boson gas final state.}
\end{figure}

The increasing trend of the cross section ratio with $\alpha$ momentum $p_{c.m.}$ is a consequence of the statistical nature of the fusion evaporation process. As the excitation energy of the compound nucleus increases, the available phase space for decay pathways involving three $\alpha$ particles becomes more favorable. This increasing trend can be explained by considering the probability distributions and partitioning of energy and angular momentum in the compound nucleus. As the alpha momentum increases, the compound nucleus has a higher probability of accessing energy states that favor the emission of multiple alpha particles, leading to an enhanced 3$\alpha$ spectrum compared to the 2$\alpha$ spectrum.

For deeply understanding this experimental results, firstly, the excess proton of $^{11}$C nucleus leads to an increased Coulomb repulsion between the protons in the collision, making the compound nucleus less stable. Secondly, the $\alpha$ particle is a tightly bound configuration in the nuclear structure. It has a high binding energy, which means it requires a large amount of energy to break apart. This stability makes the $\alpha$ particle emission a favorable decay channel for the compound nucleus, as it is less likely to undergo further reactions or break apart once emitted. Furthermore, the proton-rich projectile in a fusion evaporation reaction leads to the formation of an $\alpha$ dilute state in the compound nucleus due to the energetically favorable nature of three-body $\alpha$ correlation .

\begin{figure}
\includegraphics[width=9.2cm]{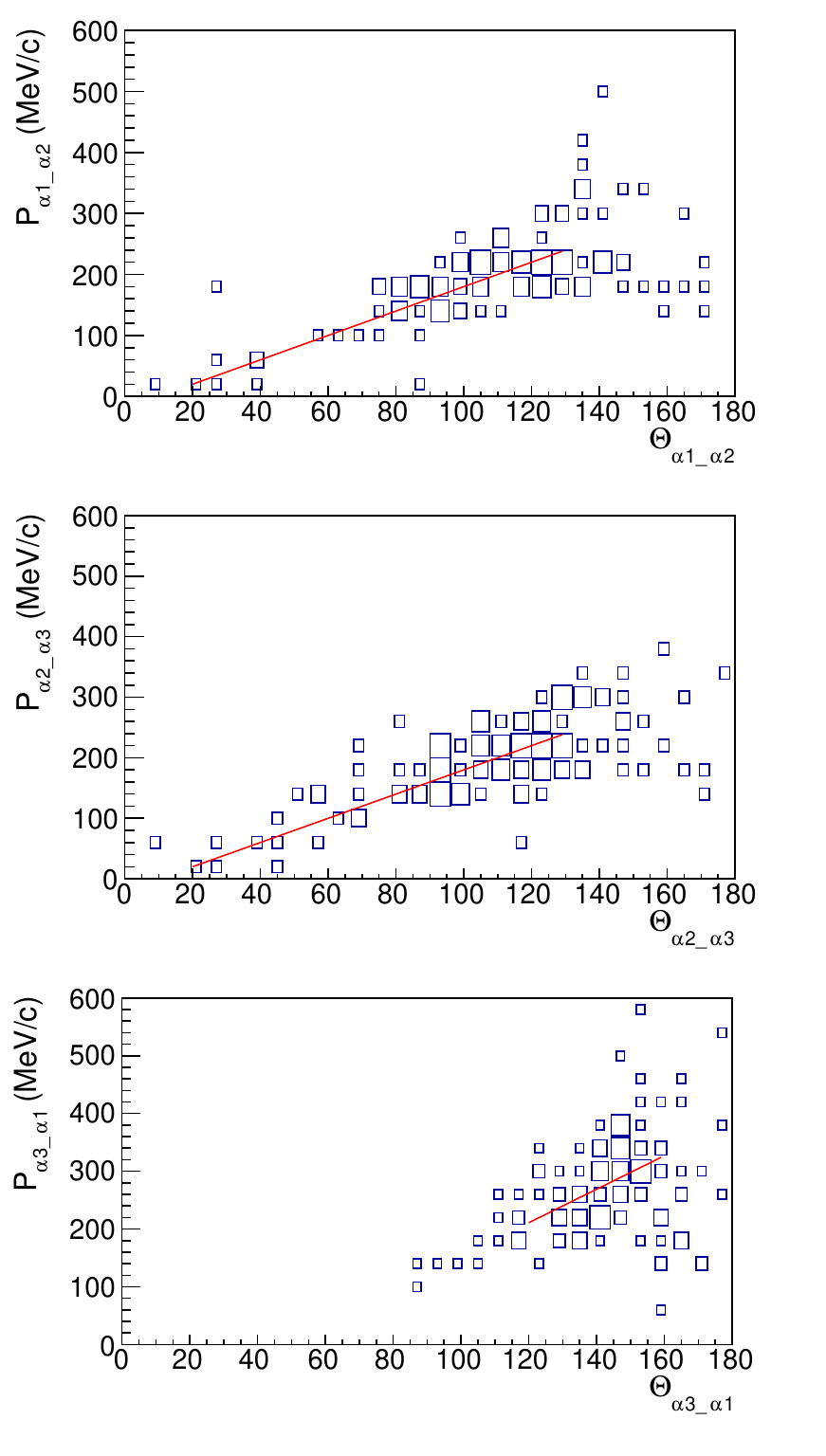}
\caption{\label{fig:pdfart}The correlation of relative momentum $P_{\alpha_{i}\_\alpha_{j}}$ and relative open emission angle $\Theta_{\alpha_{i}\_\alpha_{j}}$ for $\alpha_{1}$$\_$$\alpha_{2}$, $\alpha_{2}$$\_$$\alpha_{3}$ and $\alpha_{3}$$\_$$\alpha_{1}$ interactions.}
\end{figure}

\subsection{Evidence of $\alpha$ Boson Gas state}
The results of Boson gas model calculations are consistent with the experimental 2$\alpha$ and 3$\alpha$ spectra in fusion evaporation reactions as shown in Fig. 5, it suggests that the model accurately captures the energetics and behavior of $\alpha$ particles within the compound nucleus. This consistency indicates that the $\alpha$ cluster structure and $\alpha$ two-body and three-body interactions play a significant role in the reaction. On the other hand, if calculations using models that include additional residual particles, such as the 2$\alpha$+residue and 3$\alpha$+residue models, deviate from the experimental 2$\alpha$ and 3$\alpha$ spectra, it suggests that the involvement of these additional residule nucleus has a limited impact on the observed spectra. 

The deviation from the 2$\alpha$+residue and 3$\alpha$+residue models implies that the multi-$\alpha$ particles play a dominant role in the reaction dynamics, and their interactions are the primary driving force for the observed spectra. This deviation highlights the importance of the $\alpha$ cluster structure and the dilute state in the fusion evaporation process of $^{23}$Mg nucleus. 

\section{Discussion and conclusion}
The relative momentum $P_{\alpha_{i}\_\alpha_{j}}$ of $\alpha$$\alpha$ clusters is probably larger than that of normal or long range correlated $pp$ pair in nuclei because of the larger Coulomb force between 2$\alpha$. $P_{\alpha_{i}\_\alpha_{j}}$ reflects the correlation strength of each $\alpha_{i}\alpha_{j}$ pair in 3$\alpha$ configuration. Measurements for $P_{\alpha_{i}\_\alpha_{j}}$ can supply the significant clues for probing the properties of the initial 3$\alpha$ state. The weighted vertex of 3$\alpha$ interaction corresponds to the magnitudes of the relative momentums and relative emission angles of $\alpha_{1}$-$\alpha_{2}$, $\alpha_{2}$-$\alpha_{3}$ and $\alpha_{3}$-$\alpha_{1}$, as shown in Fig. 6. Generally, the position of weighted vertex reflects the configuration of 3$\alpha$ with an isosceles, equilateral or obtuse triangle shape. The sequence of $i$ for $\alpha_{i}$ was sorted according to the measured energies with the increase trend. 

The correlation of relative momentum $P_{\alpha_{i}\_\alpha_{j}}$ and the relative open emission angle $\Theta_{\alpha_{i}\_\alpha_{j}}$ are shown in Fig. 6. The similar magnitudes of $P_{\alpha_{1}\_\alpha_{2}}$$\approx$$P_{\alpha_{2}\_\alpha_{3}}$$\approx$216 MeV/c correspond to the approximately equal magnitudes of $\Theta_{\alpha_{1}\_\alpha_{2}}$$\approx$$\Theta_{\alpha_{2}\_\alpha_{3}}$$\approx$118$^{\circ}$, while the larger $P_{\alpha_{3}\_\alpha_{1}}$ of 305 MeV/c corresponds to the broader $\Theta_{\alpha_{3}\_\alpha_{1}}$ of 151$^{\circ}$. The two equal smaller relative momentum indicate the longer interaction range for $\alpha_{1}$ to $\alpha_{2}$ and $\alpha_{2}$ to $\alpha_{3}$, while the larger relative momentum denotes the short interaction range for $\alpha_{3}$ to $\alpha_{1}$. Therefore, the connections of $\alpha_{1}$$\_$$\alpha_{2}$, $\alpha_{2}$$\_$$\alpha_{3}$ and $\alpha_{3}$$\_$$\alpha_{1}$ make up an isosceles triangle.

In conclusion, the study of the $\alpha$ three-body force and its impact on the dilute Boson Gas state of the fusion evaporation final process of compound nucleus $^{23}$Mg has shed light on the complex dynamics involved in nuclear reactions. Through the analysis of experimental data and theoretical calculations, it has been demonstrated that the $\alpha$ three-body force plays a significant role in determining the properties of excited compound nuclei and the subsequent decay processes.

\section{Acknowledgement}
We would like to acknowledge the staff of HIRFL for the operation of the cyclotron. The author T. Wang appreciates for the financial supports from China Scholarship Council. This work has also been supported by the National Natural Science Foundation of China (No. 10175091 and No. 11305007).


\end{document}